\documentclass[runningheads]{llncs}
\usepackage[T1]{fontenc}
\usepackage{graphicx}
\usepackage{amsmath,amssymb}
\usepackage{algorithm}
\usepackage{algorithmic}
\usepackage{booktabs}
\usepackage{rotating}
\usepackage{multirow}
\usepackage{xcolor}
\usepackage{url}
\usepackage{hyperref}
\hypersetup{hidelinks}

\begin{document}

\title{LASER: Load-Aware Serving with Early-Exit for Reasoning LLMs at the Edge}

\titlerunning{LASER: Load-Aware Early-Exit Serving at the Edge}

\author{Zhiqing Tang\inst{1,2} \and 
Size Li\inst{2,1} \and 
Hanshuai Cui\inst{1,2} \and 
Zilan Huang\inst{2,1} \and 
Jianxiong Guo\inst{1,3} \and\\
Tian Wang\inst{1} \and
Yuan Wu\inst{4}\thanks{Corresponding authors.} \and
Weijia Jia\inst{1,3}\textsuperscript{$\star$}}

\authorrunning{Z. Tang et al.}

\institute{Institute of AI and Future Networks, Beijing Normal University, Zhuhai, China
\and
Faculty of Arts and Sciences, Beijing Normal University, Zhuhai, China
\and
Guangdong Key Lab of AI and Multi-Modal Data Processing, Beijing Normal-Hong Kong Baptist University, Zhuhai, China
\and
State Key Lab of IoT for Smart City, University of Macau, SAR Macau, China\\
\email{\{zhiqingtang, jianxiongguo, tianwang, jiawj\}@bnu.edu.cn, \{sizeli, hanshuaicui, zilanhuang\}@mail.bnu.edu.cn, yuanwu@um.edu.mo}}

\maketitle

\begin{abstract}
Large reasoning models (LRMs) such as DeepSeek-R1 have achieved strong performance through extended chain-of-thought (CoT) generation. However, deploying them on edge devices raises a conflict between long CoT sequences and constrained resources. Recent confidence-based early exit methods reduce CoT length for individual requests, yet they apply fixed thresholds from a single-request perspective, ignoring multi-request concurrency and load fluctuation in edge serving. To bridge this gap, we propose \underline{L}oad-\underline{A}ware \underline{S}erving with \underline{E}arly-exit for \underline{R}easoning (LASER). LASER couples two complementary designs: (1) a load-aware adaptive exit threshold that adjusts the confidence bar based on real-time system load within an empirically validated robust range, and (2) a difficulty- and load-aware reasoning budget pre-allocation that assigns compute resources by request difficulty and system capacity. We formulate the problem as a joint optimization of reasoning quality and service latency. Experiments on two reasoning models, four benchmarks, and diverse load conditions show that LASER reduces average latency by 17--38\% and improves service-level objective (SLO) satisfaction by 3--6\% over fixed-threshold baselines, at an average accuracy cost of only 1\%.

\end{abstract}

\section{Introduction}
 
The growing demand for local intelligence is driving the deployment of large language models (LLMs) to edge devices, where low latency and data privacy are essential. Large reasoning models (LRMs) such as DeepSeek-R1~\cite{deepseek_r1} and Qwen3~\cite{qwen2.5} push this frontier further by generating extended chain-of-thought (CoT) sequences to achieve strong performance on complex tasks~\cite{snell2024scaling}. Distilled variants with 1.5B--8B parameters make edge deployment technically feasible. Yet CoT sequences average 3,000--14,000 tokens per query, creating a sharp tension between the reasoning depth that accuracy demands and the limited compute budget that edge hardware can afford.
 
An overthinking problem~\cite{chen2024overthinking} further aggravates this cost. LRMs frequently produce verbose, redundant reasoning steps that consume resources without improving, and sometimes even harming, answer quality. Recent confidence-based early exit methods address this from a single-request perspective. DEER~\cite{deer} and Dynasor~\cite{dynasor} monitor model confidence during CoT generation and terminate reasoning once the model is sufficiently certain, achieving reduction in reasoning length while preserving accuracy. These training-free approaches suit edge deployment well because they require no model retraining or weight modification.
 
Existing early exit methods do not account for the serving context, applying a fixed confidence threshold to each request. At the edge, a single GPU serves requests sequentially or with minimal batching, so one request's continued reasoning directly delays those behind it. A fixed threshold may exit too early under light load or allow excessive reasoning under heavy load, hurting latency targets. Because the threshold is applied per request, the early exit decision has system-wide consequences. Therefore, a load-aware mechanism that jointly manages reasoning depth and service performance is required. To this end, two challenges must be addressed.
 
\textit{The first challenge is how to adaptively adjust the early exit threshold based on real-time system load while preserving reasoning quality.}
The confidence threshold governs the trade-off between reasoning depth and latency~\cite{deer,dynasor}: higher values usually improve accuracy but increase delay, while lower values reduce latency at the risk of premature termination~\cite{chen2024overthinking,snell2024scaling}. Unlike cloud systems with elastic scaling~\cite{distserve,llumnix}, edge servers have fixed capacity, so the threshold must adapt online. Moreover, confidence polarization means that only 2--6\% of scores fall in the intermediate range~\cite{deer}, leaving a narrow but usable tuning space for smooth load-aware adjustment.
 
\textit{The second challenge is how to pre-allocate per-request reasoning budgets considering both request difficulty and system capacity.}
Requests differ in the reasoning depth that they need: simple problems gain little from extra computation, whereas hard ones benefit more from longer reasoning~\cite{snell2024scaling}. Existing methods such as Reinforcement Learning (RL)-based length control, budget forcing, and concise-reasoning prompting largely apply uniform policies~\cite{aggarwal2025l1,arora2025efficient,chain_of_draft}, without accounting for the request difficulty and system load. In edge serving, such uniform budgets waste capacity and can hurt both throughput and accuracy. A practical allocator should therefore estimate difficulty with minimal overhead and adapt budgets to current load, complementing the adaptive threshold with a hard cap on reasoning depth~\cite{sarathi,orca}.
 
To address these challenges, we propose a Load-Aware Serving method with Early-exit for Reasoning (LASER), which elevates confidence-based early exit from a single-request optimization to a system-level scheduling mechanism for edge LLM serving. LASER couples two complementary mechanisms to jointly manage reasoning depth under varying load: a load-aware adaptive exit threshold that offers fine-grained \textit{soft} control over when to stop reasoning, and a difficulty-aware reasoning budget allocator that imposes a coarse-grained \textit{hard} cap on maximum reasoning depth. We evaluate LASER on two edge-deployed reasoning models (i.e.,  DeepSeek-R1-Distill-Qwen-7B and Qwen3-4B) across four benchmarks under five load levels using discrete-event simulation on an NVIDIA RTX 4090. The main contributions are summarized as follows.
 
\begin{enumerate}
\item \textbf{Load-aware adaptive exit threshold.} We exploit the confidence polarization phenomenon in reasoning models to establish a safe tuning space. We design a $\tanh$-based threshold function driven by exponential moving average (EMA) smoothed load signals that decreases the threshold under high load and raises it under low load.
 
\item \textbf{Difficulty- and load-aware reasoning budget allocator.} We propose a lightweight budget pre-allocation mechanism that uses prompt token count as a difficulty proxy and scales budgets proportionally with system load, providing a hard reasoning cap that complements the threshold's soft control.
 
\item \textbf{Comprehensive evaluation.} Experiments on two reasoning models show that LASER reduces average latency by 17--38\% and improves service-level objective (SLO) satisfaction by 3--6 percentage points over fixed-threshold baselines, with only 1 percentage point average accuracy loss. 
\end{enumerate}

\section{Related Work}\label{sec:related}
 
\textbf{Efficient Reasoning for Large Language Models.}  
Existing methods reduce redundant reasoning in LRMs through three main approaches. Post-training methods adapt reasoning length with variable-length CoT data or reinforcement learning~\cite{c3ot,aggarwal2025l1,arora2025efficient}, but require retraining. Prompt-based methods such as Chain-of-Draft and budget forcing~\cite{chain_of_draft} are lightweight yet may hurt quality on hard tasks. Training-free methods perform early exit during inference, for example by monitoring intermediate-answer consistency or confidence~\cite{dynasor,deer,layerskip}. However, these methods optimize each request independently and do not account for system load or queueing effects.
 
\textbf{Edge LLM Inference and Serving.}  
Prior work improves edge LLM deployment through model compression, partitioning, speculative inference, and scheduling optimizations~\cite{deepseek_r1,chen2025partition,specedge,edgeshard,fu2025serving,xu2025cadec}. LLM serving systems further improve efficiency with optimized KV-cache management, continuous batching, disaggregated execution, and runtime rescheduling~\cite{vllm,orca,sarathi,distserve,llumnix}. These approaches improve model- or system-level efficiency, but generally treat generation length as fixed. In contrast, LASER treats reasoning depth itself as a controllable serving variable and adapts it to system load.

\section{System Model and Problem Formulation}\label{sec:model}
 
\begin{figure}[t]
\centering
\includegraphics[width=0.95\textwidth]{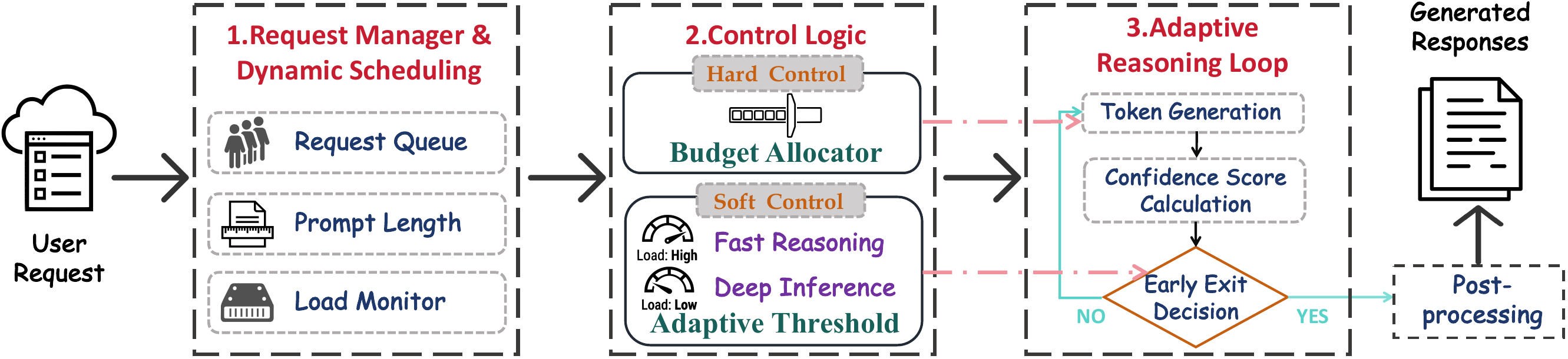}
\caption{LASER system architecture.}\label{fig:architecture}
\end{figure}

We consider an edge serving system where a single GPU-equipped server hosts a reasoning LLM to serve requests from local users, as illustrated in Fig.~\ref{fig:architecture}.

\subsection{System Model}
\textbf{Reasoning Model Generation Pattern.} 
A reasoning LLM generates output in two phases~\cite{deer}. It first performs slow thinking, enclosed in \texttt{<think>}$\ldots$\texttt{</think>} delimiters, then produces a conclusion. The slow thinking phase consists of $K_i$ reasoning chunks separated by action transition points (ATPs), typically marked by linguistic cues such as \textit{Wait} or \textit{Alternatively}.
\begin{equation}\label{eq:generation_pattern}
[\text{Prompt}] + \texttt{<think>} + T_1 + \text{ATP}_1 + T_2 + \text{ATP}_2 + \cdots + T_{K_i} + \texttt{</think>} + [\text{Conclusion}]
\end{equation}
where $T_k$ denotes the $k$-th reasoning chunk. Each ATP is a candidate early-exit point. At ATP$_k$, the model is prompted to generate a trial answer, and a confidence score $C_i^{(k)}$ is computed from the token probabilities. If $C_i^{(k)}$ exceeds the threshold $\lambda$, reasoning terminates early and the model proceeds to generate the conclusion.
 
\textbf{Cost Model.}  
For each request $r_i$, the total response latency $l_i$ consists of two parts: $l_i = w_i + t_i^{\text{inf}}$.
$w_i$ is the queueing delay and $t_i^{\text{inf}}$ is the inference time. The inference time is determined by the reasoning depth $s_i$, i.e., the number of reasoning steps executed before exit.
\begin{equation}\label{eq:inference_time}
t_i^{\text{inf}} = t_i^{\text{prefill}} + \sum_{k=1}^{s_i} \tau_k^{(i)} + t_i^{\text{conclu}}
\end{equation}
where $t_i^{\text{prefill}}$ is the prompt prefill time, $\tau_k^{(i)}$ is the time to generate the $k$-th reasoning chunk and evaluate trial answer confidence, and $t_i^{\text{conclusion}}$ is the conclusion generation time.
 
In existing early exit methods, the reasoning depth $s_i$ is determined solely by a fixed confidence threshold $\lambda$. Reasoning terminates at the first ATP$_k$ where $C_i^{(k)} > \lambda$, or when the maximum step limit is reached. A lower $\lambda$ leads to earlier exits, reducing $t_i^{\text{inf}}$ but potentially decreasing accuracy $a_i$.
 
The accuracy $a_i \in \{0,1\}$ of request $r_i$ depends on the reasoning depth $s_i$, jointly controlled by the exit threshold $\lambda_i$ and the reasoning budget $B_i$. A deeper reasoning chain increases the probability of reaching a correct answer, but with diminishing returns due to the overthinking phenomenon~\cite{chen2024overthinking}. Because $a_i$ is a stochastic function of the model's internal state, we treat it as an empirical black-box mapping $a_i(\lambda_i, B_i)$ in the optimization below.
 
\subsection{Problem Formulation}
 
Given a request stream $\mathcal{R}$, LASER jointly determines the exit threshold $\lambda_i$ and reasoning budget $B_i$ for each request to balance reasoning quality and service latency. We formulate this as the following optimization problem.
\begin{equation}\label{eq:objective}
\mathbf{P:}\quad \max_{\{\lambda_i\}, \{B_i\}} \; \frac{1}{N} \sum_{i=1}^{N} \left[ \omega \cdot a_i(\lambda_i, B_i) - (1-\omega) \cdot \frac{l_i(\lambda_i, B_i)}{d_i} \right]
\end{equation}
subject to:
\begin{equation}\label{eq:constraints}
\lambda_{\min} \leq \lambda_i \leq \lambda_{\max}, \quad
s_i \leq B_i, \quad
B_i \geq B_{\min}, \quad
\forall i
\end{equation}
where $\omega \in [0,1]$ balances accuracy and latency. The constraints require the exit threshold to remain within the empirically validated robust range $[\lambda_{\min}, \lambda_{\max}]$~\cite{deer}, the executed reasoning depth to stay within the assigned budget, and each request to receive at least the minimum reasoning budget $B_{\min}$.

\section{Algorithm Design}\label{sec:algorithm}

\begin{figure}[t]
\centering
\includegraphics[width=0.95\textwidth]{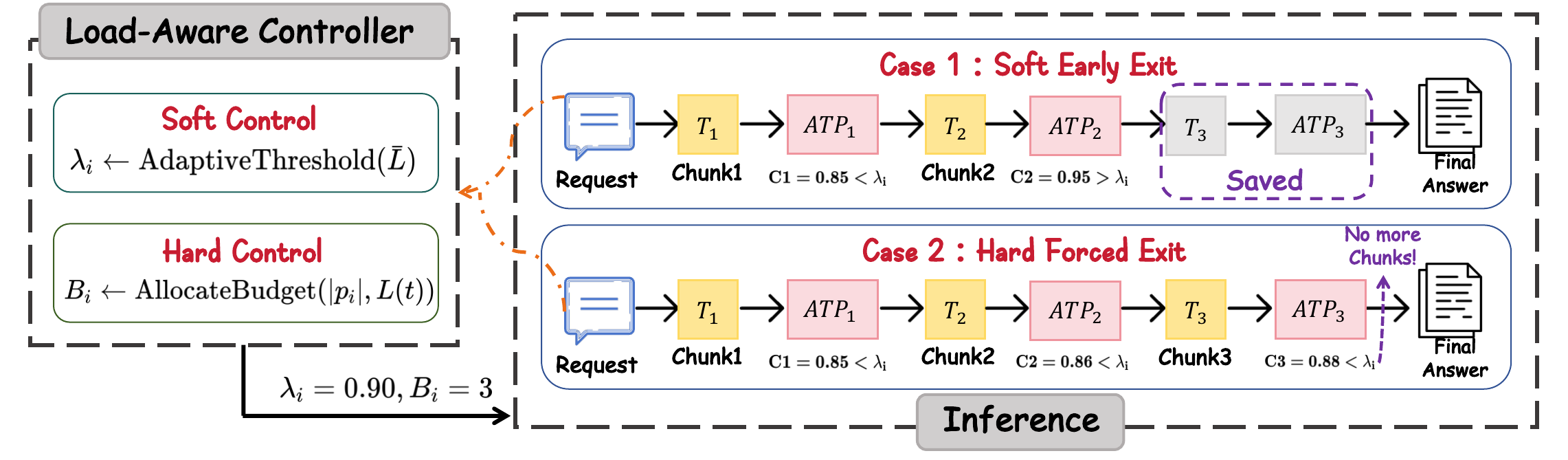}
\caption{Algorithm overview.}\label{fig:algorithm}
\end{figure}
 
\subsection{Overview}
 
LASER adds two mechanisms to the confidence-based early exit pipeline (Algorithm~\ref{alg:laser}). Before inference begins, the budget allocator assigns a per-request reasoning limit $B_i$ based on prompt difficulty and current load. During inference, the system updates the smoothed load signal at each ATP and computes a load-responsive exit threshold $\lambda_i$. Reasoning terminates when confidence exceeds $\lambda_i$ or the budget $B_i$ is exhausted. With both mechanisms disabled, LASER reduces to standard fixed-threshold early exit.
 
\begin{algorithm}[t]
\caption{LASER}\label{alg:laser}
\begin{algorithmic}[1]
\REQUIRE LRM $M$, request stream $\mathcal{R}$, base threshold $\lambda_{\text{base}}$, target load $L_{\text{target}}$, parameters $\beta, \gamma, \alpha, B_{\text{base}}$
\ENSURE Generated responses for all requests
\STATE Initialize EMA load $\bar{L} \leftarrow 0$
\FOR{each request $r_i$ with prompt $p_i$}
    \STATE $B_i \leftarrow \text{AllocateBudget}(|p_i|, L(t))$; $k \leftarrow 0$ \hfill $\triangleright$ Eq.~(\ref{eq:budget})
    \WHILE{$k < B_i$ \AND not finished}
        \STATE Generate $T_{k+1}$ until the next ATP; compute confidence $C_i^{(k+1)}$
        \STATE Update $\bar{L}$ using EMA smoothing; set $\lambda_i \leftarrow \text{AdaptiveThreshold}(\bar{L})$ by Eq.~(\ref{eq:threshold})
        \IF{$C_i^{(k+1)} > \lambda_i$}
            \STATE Terminate reasoning, generate conclusion, and mark $r_i$ as \textit{early-exited}
        \ELSE
            \STATE $k \leftarrow k + 1$
        \ENDIF
    \ENDWHILE
    \IF{$k = B_i$ \AND not finished}
        \STATE Force conclusion generation \hfill $\triangleright$ Budget exhausted
    \ENDIF
\ENDFOR
\end{algorithmic}
\end{algorithm}
 
\subsection{Load-Aware Adaptive Exit Threshold}
 
LASER replaces the fixed threshold used in existing methods~\cite{deer,dynasor} with a load-responsive function. At each confidence evaluation point, the adaptive threshold is computed as:
\begin{equation}\label{eq:threshold}
\lambda_i = \lambda_{\text{base}} - \beta \cdot \tanh\!\left(\gamma \cdot \frac{\bar{L}(t) - L_{\text{target}}}{L_{\text{target}}}\right)
\end{equation}
Here $\lambda_{\text{base}} = 0.95$ is the default threshold, $\bar{L}(t)$ is the EMA-smoothed load signal, and $L_{\text{target}}$ is the target load level. $\beta$ controls the maximum adjustment amplitude and $\gamma$ controls sensitivity to load deviations. When $\bar{L}(t) > L_{\text{target}}$, the threshold decreases to accelerate exits; when $\bar{L}(t) < L_{\text{target}}$, it increases to preserve quality. The final value is clamped to $[\lambda_{\min}, \lambda_{\max}]$.
 
We choose $\tanh$ over linear or sigmoid mappings for two reasons. First, its bounded range $(-1,1)$ helps keep $\lambda_i$ within a safe interval and limits the magnitude of load-driven adjustments. Second, its S-shaped curve produces gentle adjustment near the target load and saturates under extreme deviation, preventing overreaction to transient spikes.
 
To prevent threshold jitter from instantaneous load fluctuations, we apply EMA smoothing: $\bar{L}(t) = \alpha \cdot L(t) + (1 - \alpha) \cdot \bar{L}(t-1)$.
$L(t)$ is the current active request count and directly available as the size of the active request set in the vLLM batch processing framework.

\subsection{Difficulty- and Load-Aware Reasoning Budget Pre-Allocation}
 
While the adaptive threshold provides soft control over exit timing, LASER also assigns a hard reasoning budget $B_i$ to each request before inference begins:
\begin{equation}\label{eq:budget}
    B_i = \max\!\left\{B_{\min},\; \left\lfloor B_{\text{base}} \cdot f_{\text{diff}}(p_i) \cdot f_{\text{load}}(L(t)) \right\rfloor \right\}
\end{equation}
where $B_{\text{base}}$ is the default maximum reasoning steps and $B_{\min} \geq 2$ ensures every request receives at least minimal reasoning.

\textbf{Difficulty factor (line~3).}
The difficulty factor $f_{\text{diff}}$ uses prompt token count $|p_i|$ as a lightweight complexity proxy, linearly mapping it to a budget scaling factor clamped to $[f_{\min}, f_{\max}]$:
\begin{equation}\label{eq:difficulty}
f_{\text{diff}}(p_i) = \text{clip}\!\left(f_{\min} + \frac{|p_i| - p_{\min}}{p_{\max} - p_{\min}} \cdot (f_{\max} - f_{\min}),\; f_{\min},\; f_{\max}\right)
\end{equation}
where we set $f_{\min}=0.6$, $f_{\max}=1.5$, $p_{\min}=50$, $p_{\max}=275$ by default. Empirical evidence shows that shorter prompts exhibit higher early exit rates and require fewer reasoning steps~\cite{deer,snell2024scaling}, supporting prompt length as a difficulty proxy.
 
\textbf{Load factor (line~3).}
The load factor compresses budgets proportionally when the system is overloaded:
\begin{equation}\label{eq:load_factor}
f_{\text{load}}(L(t)) = \min\!\left(1.0,\; \frac{L_{\text{target}}}{\max(1, L(t))}\right)
\end{equation}
When $L(t) \leq L_{\text{target}}$, $f_{\text{load}} = 1.0$ and budgets are unmodified; under overload, budgets shrink proportionally. The adaptive threshold thus provides soft control over exit timing, while the budget imposes a hard cap on reasoning depth.

\subsection{Complexity Analysis}

At each ATP, LASER performs one EMA update, one $\tanh$ computation (Eq.~\ref{eq:threshold}), and one comparison, all in $O(1)$. Budget allocation (Eq.~\ref{eq:budget}) also runs in $O(1)$ per request. The total added cost for $N$ requests is $O(\sum_{i=1}^{N} s_i)$, dominated by LLM inference itself. LASER maintains only a single scalar $\bar{L}$ and a fixed set of hyperparameters, so the additional memory is $O(1)$. No extra neural network inference is required; the added latency per request is on the order of microseconds, several orders of magnitude below the per-step decoding time (10--100~ms per chunk on edge GPUs).

\section{Experiments}\label{sec:experiments}

\subsection{Experimental Setup}

\textbf{Models.}
We evaluate two reasoning LLMs from different model families: Qwen3-4B (requiring approximately 8~GB VRAM) and DeepSeek-R1-Distill-Qwen-7B (requiring approximately 14~GB VRAM). Both are edge-server-class models that fit within a single consumer-grade GPU.

\textbf{Benchmarks.}
We evaluate on four reasoning benchmarks~\cite{deer}: GSM8K~\cite{gsm8k} (1319 elementary math problems), MATH-500~\cite{math500} (500 competition-level math problems), AMC 2023~\cite{deer} (40 competition problems), and GPQA Diamond~\cite{gpqa} (198 PhD-level science questions). For reasoning quality, we report accuracy (Acc), average token count (Tok), and compression rate (CR defined as $1-\mathrm{Tok}_{\mathrm{method}}/\mathrm{Tok}_{\mathrm{vanilla}}$). For system performance, we report average latency ($\bar{l}$), P95 latency ($l_{95}$), throughput (QPS), and SLO satisfaction rate (SLO\%).

\textbf{Baselines.}
We compare LASER against four methods: (1) \textit{Vanilla}: original LRM without early exit; (2) \textit{Fixed-High}~\cite{deer}: fixed threshold $\lambda=0.95$, representing the standard early exit configuration; (3) \textit{Fixed-Low}: fixed threshold $\lambda=0.90$, representing aggressive static tuning; (4) \textit{NoThinking}: skipping the reasoning phase entirely, serving as a lower bound on reasoning quality.

\textbf{Implementation details.}
All experiments run on an NVIDIA RTX 4090 GPU (24~GB VRAM), which closely matches high-end edge accelerators such as the NVIDIA Jetson AGX Orin (64~GB unified memory). The 4B--7B parameter models fit in 24~GB VRAM, consistent with realistic edge deployment. For system-level evaluation, we employ discrete-event simulation. We first run each method on all benchmark samples to obtain per-sample inference times and accuracy, then simulate Poisson arrivals with QPS $\{0.5, 1, 2, 4, 8\}$ using a single-server FIFO queue. For burst traffic, we use a three-phase pattern of 10 minutes at 1 QPS, 10 minutes at 5 QPS, and 10 minutes at 1 QPS. Unless otherwise stated, the SLO deadline is $d = 30$ seconds and the default parameters are $\lambda_{\text{base}}=0.95$, $L_{\text{target}}=10$, $\beta=0.04$, $\gamma=1.0$, $\alpha=0.3$, $B_{\text{base}}=10$, $\lambda_{\min}=0.88$, $\lambda_{\max}=0.97$, and $B_{\min}=2$. The maximum generation length is 16384 tokens, with \texttt{think\_ratio}$\,=0.6$ for DeepSeek models and $0.8$ for Qwen3 models.

\subsection{Reasoning Quality Results}

\begin{figure}[t]
\centering
\includegraphics[width=0.8\textwidth]{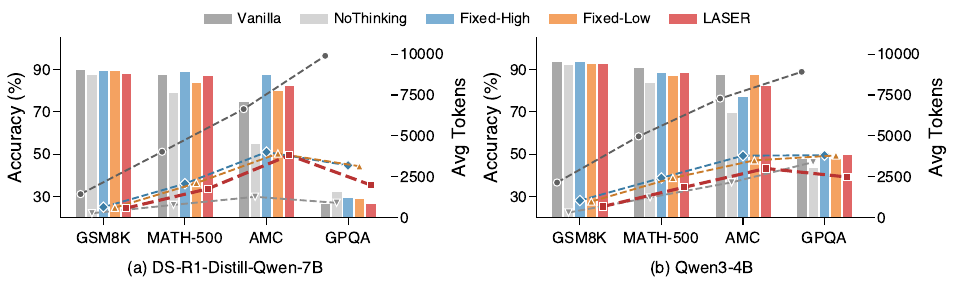}
\caption{Reasoning quality comparison across four benchmarks. Bars show accuracy (left axis) and average generated token count (right axis) for each method.}\label{fig:quality}
\end{figure}

Fig.~\ref{fig:quality} presents the reasoning quality comparison across all benchmarks for both models. LASER achieves accuracy within 2 percentage points of Fixed-High on average across both models while generating substantially fewer tokens. On DeepSeek-R1-Distill-Qwen-7B, LASER achieves 42--80\% token compression relative to Vanilla, with accuracy degradation of 1.4--5.0 percentage points compared to Fixed-High. On Qwen3-4B, LASER achieves 59--72\% compression, with accuracy within 1 percentage point of Fixed-High on GSM8K and MATH-500 and even outperforming Fixed-High on AMC by 5.0 pp. 

\subsection{System Performance Under Varying Loads}

Table~\ref{tab:qps} presents system-level performance under varying QPS levels on MATH-500 for both models. Under low load (QPS=0.5), queueing effects are minimal and LASER's advantage is most visible in raw latency reduction (36--38\%). As load increases, all methods experience higher latency due to queueing, but LASER consistently outperforms fixed-threshold baselines. At QPS=2, LASER reduces average latency by 20--23\% compared to Fixed-High. Under high load (QPS=8), the reduction remains 17--20\%. Throughput improvement is consistent across load levels, with LASER achieving 0.32--0.34 QPS versus 0.28--0.31 for Fixed-High, a 6--17\% gain. SLO satisfaction under LASER exceeds Fixed-High by 4--6 percentage points under medium and high loads.

\begin{table}[t]
\caption{System performance on MATH-500 under varying QPS. $\bar{l}$: avg latency (s), $l_{95}$: P95 latency (s), Tput: throughput (QPS), SLO: SLO satisfaction (\%).}\label{tab:qps}
\centering
\tiny
\begin{tabular}{ll rrrr rrrr rrrr}
\toprule
& & \multicolumn{4}{c}{\textbf{QPS=0.5}} & \multicolumn{4}{c}{\textbf{QPS=2.0}} & \multicolumn{4}{c}{\textbf{QPS=8.0}} \\
\cmidrule(lr){3-6} \cmidrule(lr){7-10} \cmidrule(lr){11-14}
\textbf{Model} & \textbf{Method} & $\bar{l}$ & $l_{95}$ & Tput & SLO & $\bar{l}$ & $l_{95}$ & Tput & SLO & $\bar{l}$ & $l_{95}$ & Tput & SLO \\
\midrule
\multirow{5}{*}{{\scriptsize DS-R1-7B}}
& Vanilla     & 248.5 & 484.5 & 0.17 & 8.0  & 329.5 & 640.9 & 0.17 & 7.2  & 350.3 & 680.9 & 0.17 & 7.2 \\
& NoThinking  & 8.6   & 27.4  & 0.54 & 100  & 51.4  & 85.1  & 0.86 & 26.4 & 72.2  & 124.7 & 0.86 & 11.2 \\
& Fixed-High  & 93.9  & 219.1 & 0.28 & 13.6 & 167.9 & 369.1 & 0.29 & 13.6 & 188.7 & 409.0 & 0.29 & 12.0 \\
& Fixed-Low   & 89.5  & 156.0 & 0.33 & 8.8  & 166.3 & 309.1 & 0.33 & 7.2  & 187.1 & 349.0 & 0.33 & 7.2 \\
& LASER       & 58.4  & 149.2 & 0.33 & 17.6 & 130.0 & 297.0 & 0.34 & 17.6 & 150.8 & 336.9 & 0.34 & 16.8 \\
\midrule
\multirow{5}{*}{{\scriptsize Qwen3-4B}}
& Vanilla     & 317.9 & 591.1 & 0.15 & 7.2  & 399.6 & 751.1 & 0.15 & 4.8  & 420.4 & 791.2 & 0.15 & 4.8 \\
& NoThinking  & 14.4  & 46.5  & 0.46 & 82.4 & 65.4  & 147.1 & 0.60 & 14.4 & 86.2  & 186.4 & 0.60 & 8.8 \\
& Fixed-High  & 76.9  & 178.8 & 0.30 & 16.0 & 157.4 & 334.1 & 0.31 & 9.6  & 178.2 & 374.0 & 0.31 & 8.8 \\
& Fixed-Low   & 92.9  & 204.3 & 0.28 & 14.4 & 173.2 & 361.4 & 0.29 & 12.0 & 194.0 & 401.5 & 0.29 & 9.6 \\
& LASER       & 49.2  & 155.3 & 0.32 & 61.6 & 126.6 & 308.4 & 0.33 & 15.2 & 147.4 & 348.5 & 0.33 & 13.6 \\
\bottomrule
\end{tabular}
\end{table}

\subsection{Burst Traffic Response}

We evaluate LASER's response to sudden traffic spikes using a three-phase load pattern on MATH-500 (1$\to$5$\to$1 QPS, equal-duration phases). Table~\ref{tab:burst} reports the results. On DS-R1-Distill-Qwen-7B, LASER reduces average latency from 175.4~s (Fixed-High) to 137.5~s (21.6\% reduction) and improves SLO satisfaction from 13.6\% to 17.6\%. On Qwen3-4B, LASER achieves 133.7~s versus 163.9~s for Fixed-High (18.4\% reduction), with SLO satisfaction improving from 12.0\% to 17.6\%. LASER detects the load spike through its EMA-smoothed signal and lowers the threshold to accelerate exits during the burst phase, then gradually restores quality as load returns to normal.

\begin{table}[t]
\caption{Burst traffic performance on MATH-500 (three-phase: 1$\to$5$\to$1 QPS).}\label{tab:burst}
\centering
\tiny
\begin{tabular}{lrrrr lrrrr}
\toprule
\multicolumn{5}{c}{\textbf{DS-R1-7B}} & \multicolumn{5}{c}{\textbf{Qwen3-4B}} \\
\cmidrule(lr){1-5} \cmidrule(lr){6-10}
\textbf{Method} & $\bar{l}$ \textbf{(s)} & $l_{95}$ \textbf{(s)} & \textbf{Tput} & \textbf{SLO\%} & \textbf{Method} & $\bar{l}$ \textbf{(s)} & $l_{95}$ \textbf{(s)} & \textbf{Tput} & \textbf{SLO\%} \\
\midrule
Vanilla    & 335.6 & 657.1 & 0.17 & 7.2  & Vanilla    & 405.0 & 767.3 & 0.15 & 5.6 \\
NoThinking & 60.7  & 105.2 & 0.83 & 26.4 & NoThinking & 73.8  & 165.9 & 0.59 & 14.4 \\
Fixed-High & 175.4 & 387.3 & 0.29 & 13.6 & Fixed-High & 163.9 & 351.5 & 0.31 & 12.0 \\
Fixed-Low  & 173.9 & 326.8 & 0.33 & 8.8  & Fixed-Low  & 180.0 & 379.0 & 0.29 & 13.6 \\
LASER      & 137.5 & 315.4 & 0.34 & 17.6 & LASER      & 133.7 & 326.4 & 0.33 & 17.6 \\
\bottomrule
\end{tabular}
\end{table}

\subsection{Ablation Study}

\begin{figure}[t]
\centering
\includegraphics[width=\textwidth]{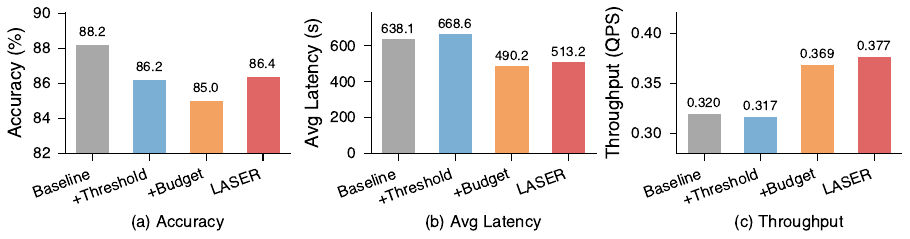}
\caption{Ablation study on MATH-500 with DS-R1-Distill-Qwen-7B at QPS=2. Budget allocation is the primary driver of latency reduction, while the adaptive threshold contributes accuracy recovery when combined.}\label{fig:ablation}
\end{figure}

Fig.~\ref{fig:ablation} presents the ablation study on DS-R1-Distill-Qwen-7B and MATH-500 under medium load (QPS=2), isolating the contribution of each component. Budget allocation is the primary driver of latency reduction, decreasing average latency by 23.2\% compared to the fixed-threshold baseline. The adaptive threshold alone does not reduce latency because without budget control it merely shifts the exit point within the reasoning chain without capping the maximum depth. 
When combined, the two mechanisms enable LASER to achieve the best throughput while recovering 1.4 percentage points of accuracy compared with the budget-only variant. The adaptive threshold preserves reasoning quality by allowing deeper reasoning under light load, complementing the budget allocator.

\subsection{Parameter Sensitivity}

Table~\ref{tab:sens} reports the sensitivity of LASER to its key hyperparameters on MATH-500 at QPS=2. LASER shows robust performance across a wide range of parameter values. For $\beta$, accuracy remains within 83--89\% across the sweep and latency varies smoothly. For $L_{\text{target}}$, higher values tend to preserve accuracy while lower values favor throughput. Moderate EMA smoothing ($\alpha \in [0.2, 0.5]$) achieves a good balance between responsiveness and stability. The $\tanh$ function naturally bounds the adjustment, providing self-correcting behavior that reduces the need for per-scenario tuning in edge deployments.

\begin{table}[t]
    \caption{Parameter sensitivity on MATH-500 at QPS=2.}\label{tab:sens}
    \centering
    \tiny
    \begin{tabular}{ll rrrrr rrrrr rrrrr}
    \toprule
    & & \multicolumn{5}{c}{$\beta$} & \multicolumn{5}{c}{$L_{\text{target}}$} & \multicolumn{5}{c}{$\alpha$} \\
    \cmidrule(lr){3-7} \cmidrule(lr){8-12} \cmidrule(lr){13-17}
    \textbf{Model} & \textbf{Metric} & 0.02 & 0.03 & 0.04 & 0.05 & 0.06 & 5 & 8 & 10 & 15 & 20 & 0.1 & 0.2 & 0.3 & 0.5 & 0.7 \\
    \midrule
    \multirow{4}{*}{\textbf{DS-R1}}
    & \textbf{Acc}  & 82.5 & 84.1 & 85.7 & 84.1 & 84.1 & 82.5 & 85.7 & 81.0 & 85.7 & 87.3 & 88.9 & 87.3 & 87.3 & 82.5 & 84.1 \\
    & $\bar{l}$      & 98.6 & 45.4 & 107.3 & 78.5 & 59.6 & 81.4 & 78.1 & 81.2 & 76.8 & 58.5 & 86.7 & 69.2 & 75.9 & 67.9 & 89.4 \\
    & \textbf{Tput} & 0.33 & 0.50 & 0.30 & 0.36 & 0.46 & 0.37 & 0.36 & 0.35 & 0.36 & 0.46 & 0.35 & 0.41 & 0.37 & 0.43 & 0.33 \\
    & \textbf{SLO\%} & 15.9 & 38.1 & 14.3 & 28.6 & 14.3 & 22.2 & 30.2 & 17.5 & 30.2 & 33.3 & 14.3 & 27.0 & 30.2 & 27.0 & 15.9 \\
    \midrule
    \multirow{4}{*}{\textbf{Qwen3}}
    & \textbf{Acc}  & 92.1 & 87.3 & 88.9 & 85.7 & 88.9 & 87.3 & 88.9 & 88.9 & 87.3 & 92.1 & 93.7 & 88.9 & 84.1 & 87.3 & 87.3 \\
    & $\bar{l}$      & 53.9 & 62.6 & 72.7 & 70.2 & 88.5 & 69.5 & 76.5 & 49.0 & 78.4 & 47.5 & 54.3 & 72.2 & 73.9 & 58.5 & 76.8 \\
    & \textbf{Tput} & 0.47 & 0.43 & 0.41 & 0.42 & 0.33 & 0.40 & 0.41 & 0.51 & 0.36 & 0.51 & 0.39 & 0.38 & 0.41 & 0.43 & 0.37 \\
    & \textbf{SLO\%} & 27.0 & 19.1 & 17.5 & 14.3 & 19.1 & 17.5 & 17.5 & 30.2 & 28.6 & 28.6 & 28.6 & 28.6 & 17.5 & 27.0 & 17.5 \\
    \bottomrule
    \end{tabular}
    \end{table}

\section{Conclusion}\label{sec:conclusion}

This paper presents LASER, a load-aware early exit approach for reasoning model serving at the edge. We have identified the gap between single-request reasoning optimization and system-level edge serving, and addressed it with two complementary mechanisms: an adaptive exit threshold driven by real-time system load and a difficulty- and load-aware reasoning budget allocator. The budget allocator serves as the primary driver of latency reduction by capping per-request reasoning depth according to difficulty and load, while the adaptive threshold exploits the confidence polarization phenomenon to preserve reasoning quality within a validated safe range. Experiments on two models, four benchmarks, and diverse load conditions show that LASER reduces average latency by 17--38\% and improves SLO satisfaction by 3--6 percentage points over fixed-threshold baselines, at an average accuracy cost of only 1\%. In future work, we will extend LASER to multi-device edge clusters with heterogeneous GPUs and integrate reasoning depth control with model selection routing.

\begin{credits}
\subsubsection{\ackname}
This work was supported in part by the National Natural Science Foundation of China (NSFC) under Grant 62302048, Grant 62272050, and Grant U25A20436; in part by the Science and Technology Development Fund of Macau SAR under Grants 0028/2025/AFJ and 0021/2025/RIA1; in part by Guangdong Higher Education Association under Grant 24GQN97; in part by the Guangdong Provincial Higher Education Institutions under Grant 2024KTSCX219; and in part by Beijing Normal University at Zhuhai Education Reform Project under Grant jx2025037.

\subsubsection{\discintname}
The authors have no competing interests to declare that are relevant to the content of this article.
\end{credits}

\bibliographystyle{splncs04}
\bibliography{references}

@article{deepseek_r1,
  title={Deepseek-r1: Incentivizing reasoning capability in llms via reinforcement learning},
  author={Guo, Daya and Yang, Dejian and Zhang, Haowei and Song, Junxiao and Wang, Peiyi and Zhu, Qihao and Xu, Runxin and Zhang, Ruoyu and Ma, Shirong and Bi, Xiao and others},
  journal={arXiv preprint arXiv:2501.12948},
  year={2025}
}

@article{qwen2.5,
  title={Qwen2. 5-coder technical report},
  author={Hui, Binyuan and Yang, Jian and Cui, Zeyu and Yang, Jiaxi and Liu, Dayiheng and Zhang, Lei and Liu, Tianyu and Zhang, Jiajun and Yu, Bowen and Lu, Keming and others},
  journal={arXiv preprint arXiv:2409.12186},
  year={2024}
}

@article{snell2024scaling,
  title={Scaling llm test-time compute optimally can be more effective than scaling model parameters},
  author={Snell, Charlie and Lee, Jaehoon and Xu, Kelvin and Kumar, Aviral},
  journal={arXiv preprint arXiv:2408.03314},
  year={2024}
}

@article{chen2024overthinking,
  title={Do not think that much for 2+ 3=? on the overthinking of o1-like llms},
  author={Chen, Xingyu and Xu, Jiahao and Liang, Tian and He, Zhiwei and Pang, Jianhui and Yu, Dian and Song, Linfeng and Liu, Qiuzhi and Zhou, Mengfei and Zhang, Zhuosheng and others},
  journal={arXiv preprint arXiv:2412.21187},
  year={2024}
}

@article{deer,
  title={Dynamic early exit in reasoning models},
  author={Yang, Chenxu and Si, Qingyi and Duan, Yongjie and Zhu, Zheliang and Zhu, Chenyu and Li, Qiaowei and Chen, Minghui and Lin, Zheng and Wang, Weiping},
  journal={arXiv preprint arXiv:2504.15895},
  year={2025}
}

@inproceedings{c3ot,
  title={C3ot: Generating shorter chain-of-thought without compromising effectiveness},
  author={Kang, Yu and Sun, Xianghui and Chen, Liangyu and Zou, Wei},
  booktitle={Proceedings of the AAAI Conference on Artificial Intelligence},
  volume={39},
  number={23},
  pages={24312--24320},
  year={2025}
}

@article{aggarwal2025l1,
  title={L1: Controlling how long a reasoning model thinks with reinforcement learning},
  author={Aggarwal, Pranjal and Welleck, Sean},
  journal={arXiv preprint arXiv:2503.04697},
  year={2025}
}

@inproceedings{arora2025efficient,
  title={Training Language Models to Reason Efficiently},
  author={Arora, Daman and Zanette, Andrea},
  booktitle={The Thirty-ninth Annual Conference on Neural Information Processing Systems (NeurIPS)},
  year={2025}
}

@article{chain_of_draft,
  title={Chain of draft: Thinking faster by writing less},
  author={Xu, Silei and Xie, Wenhao and Zhao, Lingxiao and He, Pengcheng},
  journal={arXiv preprint arXiv:2502.18600},
  year={2025}
}

@inproceedings{dynasor,
  title={Reasoning without self-doubt: More efficient chain-of-thought through certainty probing},
  author={Fu, Yichao and Chen, Junda and Zhuang, Yonghao and Fu, Zheyu and Stoica, Ion and Zhang, Hao},
  booktitle={ICLR 2025 Workshop on Foundation Models in the Wild},
  year={2025}
}

@inproceedings{layerskip,
  title={Layerskip: Enabling early exit inference and self-speculative decoding},
  author={Elhoushi, Mostafa and Shrivastava, Akshat and Liskovich, Diana and Hosmer, Basil and Wasti, Bram and Lai, Liangzhen and Mahmoud, Anas and Acun, Bilge and Agarwal, Saurabh and Roman, Ahmed and others},
  booktitle={Proceedings of the 62nd Annual Meeting of the Association for Computational Linguistics (Volume 1: Long Papers)},
  pages={12622--12642},
  year={2024}
}

@article{specedge,
  title={Specedge: Scalable edge-assisted serving framework for interactive llms},
  author={Park, Jinwoo and Cho, Seunggeun and Han, Dongsu},
  journal={arXiv preprint arXiv:2505.17052},
  year={2025}
}

@article{edgeshard,
  title={Edgeshard: Efficient llm inference via collaborative edge computing},
  author={Zhang, Mingjin and Shen, Xiaoming and Cao, Jiannong and Cui, Zeyang and Jiang, Shan},
  journal={IEEE Internet of Things Journal},
  volume={12},
  number={10},
  pages={13119--13131},
  year={2024},
  publisher={IEEE}
}

@article{fu2025serving,
  title={Serving transformer models via joint requst scheduling and batching in the network edge},
  author={Fu, Boqian and Chen, Fahao and Li, Peng and Zeng, Deze},
  journal={IEEE Transactions on Sustainable Computing},
  volume={10},
  number={4},
  pages={678--689},
  year={2025},
  publisher={IEEE}
}

@article{xu2025cadec,
  title={Cadec: a combinatorial auction for dynamic distributed dnn inference scheduling in edge-cloud networks},
  author={Xu, Xiaolong and Hu, Yuhao and Cui, Guangming and Qi, Lianyong and Dou, Wanchun and Cai, Zhipeng},
  journal={IEEE transactions on Mobile computing},
  year={2025},
  publisher={IEEE}
}

@inproceedings{chen2025partition,
  title={Latency-Optimal and Memory-Aware Model Partitioning for Cooperative Inference at the Edge},
  author={Chen, Quan and Gao, Hong and Yi, Ming and Li, Jing and Cheng, Lianglun and Li, Yingshu},
  booktitle={International Conference on Wireless Artificial Intelligent Computing Systems and Applications (WASA)},
  pages={25--37},
  year={2025}
}

@inproceedings{vllm,
  title={Efficient memory management for large language model serving with pagedattention},
  author={Kwon, Woosuk and Li, Zhuohan and Zhuang, Siyuan and Sheng, Ying and Zheng, Lianmin and Yu, Cody Hao and Gonzalez, Joseph and Zhang, Hao and Stoica, Ion},
  booktitle={Proceedings of the 29th symposium on operating systems principles (SOSP)},
  pages={611--626},
  year={2023}
}

@inproceedings{orca,
  title={Orca: A distributed serving system for $\{$Transformer-Based$\}$ generative models},
  author={Yu, Gyeong-In and Jeong, Joo Seong and Kim, Geon-Woo and Kim, Soojeong and Chun, Byung-Gon},
  booktitle={16th USENIX Symposium on Operating Systems Design and Implementation (OSDI)},
  pages={521--538},
  year={2022}
}

@inproceedings{sarathi,
  title={Taming throughput-latency tradeoff in LLM inference with sarathi-serve},
  author={Agrawal, Amey and Kedia, Nitin and Panwar, Ashish and Mohan, Jayashree and Kwatra, Nipun and Gulavani, Bhargav S and Tumanov, Alexey and Ramjee, Ramachandran},
  booktitle={18th USENIX Symposium on Operating Systems Design and Implementation (OSDI)},
  pages={117--134},
  year={2024}
}

@inproceedings{distserve,
  title={DistServe: Disaggregating prefill and decoding for goodput-optimized large language model serving},
  author={Zhong, Yinmin and Liu, Shengyu and Chen, Junda and Hu, Jianbo and Zhu, Yibo and Liu, Xuanzhe and Jin, Xin and Zhang, Hao},
  booktitle={18th USENIX Symposium on Operating Systems Design and Implementation (OSDI)},
  pages={193--210},
  year={2024}
}

@inproceedings{llumnix,
  title={Llumnix: Dynamic scheduling for large language model serving},
  author={Sun, Biao and Huang, Ziming and Zhao, Hanyu and Xiao, Wencong and Zhang, Xinyi and Li, Yong and Lin, Wei},
  booktitle={18th USENIX Symposium on Operating Systems Design and Implementation (OSDI)},
  pages={173--191},
  year={2024}
}

@article{gsm8k,
  title={Training verifiers to solve math word problems},
  author={Cobbe, Karl and Kosaraju, Vineet and Bavarian, Mohammad and Chen, Mark and Jun, Heewoo and Kaiser, Lukasz and Plappert, Matthias and Tworek, Jerry and Hilton, Jacob and Nakano, Reiichiro and others},
  journal={arXiv preprint arXiv:2110.14168},
  year={2021}
}

@inproceedings{math500,
  title={Measuring Mathematical Problem Solving With the MATH Dataset},
  author={Hendrycks, Dan and Burns, Collin and Kadavath, Saurav and Arora, Akul and Basart, Steven and Tang, Eric and Song, Dawn and Steinhardt, Jacob},
  booktitle={The Thirty-fifth Conference on Neural Information Processing Systems (NeurIPS)},
  year={2021}
}

@inproceedings{gpqa,
  title={Gpqa: A graduate-level google-proof q\&a benchmark},
  author={Rein, David and Hou, Betty Li and Stickland, Asa Cooper and Petty, Jackson and Pang, Richard Yuanzhe and Dirani, Julien and Michael, Julian and Bowman, Samuel R},
  booktitle={First conference on language modeling (COLM)},
  year={2024}
}

\end{document}